# Carbon Free High Loading Silicon Anodes Enabled by Sulfide Solid Electrolytes for Robust All Solid-State Batteries

Darren H. S. Tan,[1] Yu-Ting Chen,[1] Hedi Yang,[1] Wurigumula Bao,[1] Bhagath Sreenarayanan,[1] Jean-Marie Doux,[1] Weikang Li,[1] Bingyu Lu,[1] So-Yeon Ham,[1] Baharak Sayahpour,[1] Jonathan Scharf,[1] Erik A. Wu,[1] Grayson Deysher,[1] Hyea Eun Han,[2] Hoe Jin Hah,[2] Hyeri Jeong,[2] Zheng Chen[1,3,4,*] and Ying Shirley Meng[1,4,*]

[1]Department of NanoEngineering, University of California San Diego, La Jolla, California 92093, United States.

[2]LG Energy Solution, Ltd., LG Science Park, Magokjungang 10-ro, Gangseo-gu, Seoul, 07796, Korea

[3]Program of Chemical Engineering, University of California San Diego, La Jolla, California 92093, United States.

[4]Sustainable Power & Energy Center (SPEC), University of California San Diego, La Jolla, California 92093, United States.

*Correspondence to: shmeng@ucsd.edu, zhengchen@eng.ucsd.edu

**Abstract:** The development of silicon anodes to replace conventional graphite in efforts to increase energy densities of lithium-ion batteries has been largely impeded by poor interfacial stability against liquid electrolytes. Here, stable operation of 99.9 weight% micro-Si (μSi) anode is enabled by utilizing the interface passivating properties of sulfide based solid-electrolytes. Bulk to surface characterization, as well as quantification of interfacial components showed that such an approach eliminates continuous interfacial growth and irreversible lithium losses. In μSi ∥ layered-oxide full cells, high current densities at room temperature (5 mA cm$^{-2}$), wide operating temperature (-20°C to 80°C) and high loadings (>11 mAh cm$^{-2}$) were demonstrated for both charge and discharge operations. The promising battery performance can be attributed to both the desirable interfacial property between μSi and sulfide electrolytes, as well as the unique chemo-mechanical behavior of the Li-Si alloys.

**One Sentence Summary:** Solid electrolytes enable microparticle Si anodes, free of carbon additives and excessive binders, for stable high performance all solid-state batteries.

**Introduction:**

Silicon (Si), with a specific capacity exceeding 3500 mAh g$^{-1}$, has emerged as a promising alternative to graphite-based anodes (with specific capacity of around 370 mAh g$^{-1}$) in order to increase energy densities of lithium ion batteries (LIBs), to serve various energy storage applications such as electric vehicles (EVs) and portable devices.(*1, 2*) Beyond being the second most abundant element in the Earth's crust, it is also environmentally benign and exhibits electrochemical potentials close to graphite (0.3 V vs Li/Li$^+$).(*2*) Unfortunately, commercialization of Si anodes is hindered by its poor cycle and calendar life resulting from continuous solid electrolyte interphase (SEI) growth between the highly reactive Li-Si alloy and organic liquid electrolytes used in LIBs. This is further exacerbated by its large volumetric expansion (>300%) during lithiation, and loss of Li$^+$ inventory due to irreversible trapped Li-Si alloy enclosed within the SEI formed.(*3*) Potentially effective solutions need to address the fundamental failure mechanisms at the Si electrode-electrolyte interface that cause poor cycle and calendar life. To realize high energy density, binder and carbon conductive additive ratios also need to be kept at a minimum.

Current efforts to mitigate capacity fade include the use of sophisticated Si nanostructures in combination with carbon composites and robust binder matrix to mitigate pulverization.(*1, 3-6*) Liquid electrolyte modifications, including the use of cyclic ethers, fluorinated additives or other ionic liquids additives that stabilize the SEI have also been explored.(*5, 7*) While improvements have been reported in numerous half-cell studies, the uncontrolled amounts of lithium excess used were difficult to determine, which makes it challenging to evaluate the strategies proposed. Amongst reports that demonstrated stable cycling in full cells(*5, 8-20*), most adopt composites containing between 60 to 80 wt% Si, with carbon additives and polymeric binders typically making up the rest of the electrode. Additionally, most reported full cell performances are limited to 100 cycles, apart from a few that demonstrated longer cycle life using various pre-lithiation strategies to compensate for Li$^+$ inventory losses.(*9, 14, 18*) While pre-lithiation is believed to be effective to extend cycle life, the ideal Si anode should be composed of pristine µSi particles that do not require further treatment, reaping the benefits of low costs, ambient air-stability and environmentally benign properties. To realize this potential, it is vital to address the two key challenges of µSi anodes: a) achieving high µSi loading with minimal incorporation of carbon and binder, b) stabilizing interfacial growth originating from volume expansion and Li$^+$ consumption.

Recognizing that the Si stability problems arise mainly from the liquid electrolyte interface, the use of solid-state electrolytes (SSEs) in an all solid-state battery (ASSB) cell configuration is a promising alternative approach, due to its ability to form a stable SEI(*21*). While previous studies have reported the use of thin (sub-micron) film type Si in ASSBs, without use of carbon or binder(*11, 22, 23*), none have explored use of bulk type µSi for fabricating high-loading anodes to date. Most ASSB reports have focused instead on the use of metallic Li, in efforts to maximize cell energy densities.(*24, 25*) However, small critical current densities of metallic Li anodes often dictate the need for operation under elevated temperatures especially during cell charging. The shortcomings of bulk µSi anodes in liquid electrolytes and metallic Li anodes in ASSBs presents us with an opportunity to combine bulk µSi anodes with SSEs, directly addressing the fundamental problems of poor interface stability of Si anodes in liquid electrolytes and limited critical current of metallic Li in ASSBs at room temperature.

In this work, µSi electrode consisting of 99.9 wt% Si is used in a Li$_6$PS$_5$Cl argyrodite SSE system to enable high current densities as well as long cycle stability of a µSi||SSE||lithium nickel cobalt

manganese oxide (NCM811) full cell. This is achieved by adopting µSi-based electrodes prepared using 0.1wt% PVDF binder and no additional carbon additives (**Fig 1**). As bulk µSi already exhibits a sufficient electronic conductivity of about $3 \times 10^{-5}$ S cm$^{-1}$ (**Fig S1**), comparable to most common cathode materials ($\sim 10^{-6}$ to $10^{-4}$ S cm$^{-1}$), carbon additives are no longer necessary.(*26-28*) Moreover, carbon has also been found to be detrimental to the stability of SSEs, as it increases the kinetics of SSE decomposition at electrode interfaces.(*29, 30*) Unlike its liquid counterparts, the SSE does not permeate through the porous µSi electrode, and the interfacial contact area between the SSE and the µSi electrode is significantly reduced to a 2D plane. As a result, during cell cycling, the passivating nature of the SEI formed dramatically reduces both Li$^+$ consumption as well as trapped Li-Si typically observed in liquid electrolyte-based cells. During lithiation of µSi, Li-Si formation can propagate throughout the electrode, benefiting from the direct contact between Li-Si and Si particles (**Fig 1**). This process was found to be highly reversible at relatively high current densities up to 5 mA cm$^{-2}$, able to operate between -20°C to 80°C, as well as deliver high areal capacities of up to 11 mAh cm$^{-2}$ with specific capacity of 2890 mAh g$^{-1}$. Without any excess lithium, the µSi-NCM811 full cell was found to deliver a capacity retention of 80% after 500 cycles, demonstrating the overall robustness of µSi anodes enabled by ASSBs.

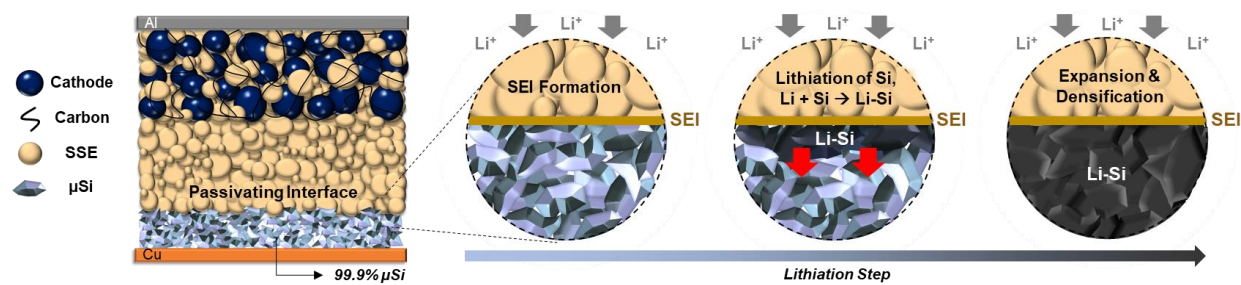

***Fig. 1.*** *Schematic of 99.9 wt% µSi electrode in an ASSB full cell. 1) During lithiation, a passivating SEI is formed between the µSi electrode and the SSE, followed by lithiation of µSi particles near the interface. 2) The highly reactive Li-Si then reacts with Si particles within its vicinity. 3) The reaction propagates throughout the electrode, forming a densified Li-Si layer.*

**Results:**
*Interface Characterization.* To demonstrate the importance of eliminating carbon in the anode, as well as the passivating nature of the Si-SSE interface, the SEI products as well as the extent of SSE decomposition were characterized and quantified with and without presence of carbon additives. While Li metal is typically used as the counter electrode in liquid electrolyte studies, its low critical current density in ASSBs make it unsuitable in our system.(*31, 32*) Thus, NCM811 was instead chosen as the counter electrode, allowing direct evaluation of µSi in a full cell. To prepare the samples, two µSi-SSE-NCM811 cells were assembled (with and without 20 wt.% carbon additives). **Figure 2a** shows the voltage profiles of both cells during the first lithiation. The cell without carbon shows an initial voltage plateau around 3.5 V, typical of a µSi||NCM811 full cell. However, the cell with 20 wt.% carbon shows a stark difference, with a lower initial plateau at 2.5 V, indicating electrochemical decomposition of the SSE before reaching the lithiation potential above 3.5 V. Previous studies have found that sulfide SSEs reduce at potentials of around 1 V *vs* Li/Li$^+$, which agrees with the observations of the different initial voltage plateau when carbon is used.(*29*) To characterize the SSE decomposition, **Fig 2b** compares the diffraction patterns of the pristine Si-SSE, lithiated Si-SSE and lithiated Si-SSE-carbon interfaces. The lithiated Si-SSE sample retained the crystalline structure of the SSE as well as the unreacted Si, with some signals of amorphous Li-Si manifesting as a hump at around 20°. While some SEI is expected, the low amount formed as the interface is most likely not detectable using this bulk technique. However, in the cell where carbon is used, most of the pristine SSE's diffraction signals are no longer present, indicating severe electrolyte decomposition. During this process, nanocrystalline $Li_2S$ forms as a major electrolyte decomposition product and is observed as broad peaks appearing at $2\theta$ angles of around 26°, 45° and 52°.

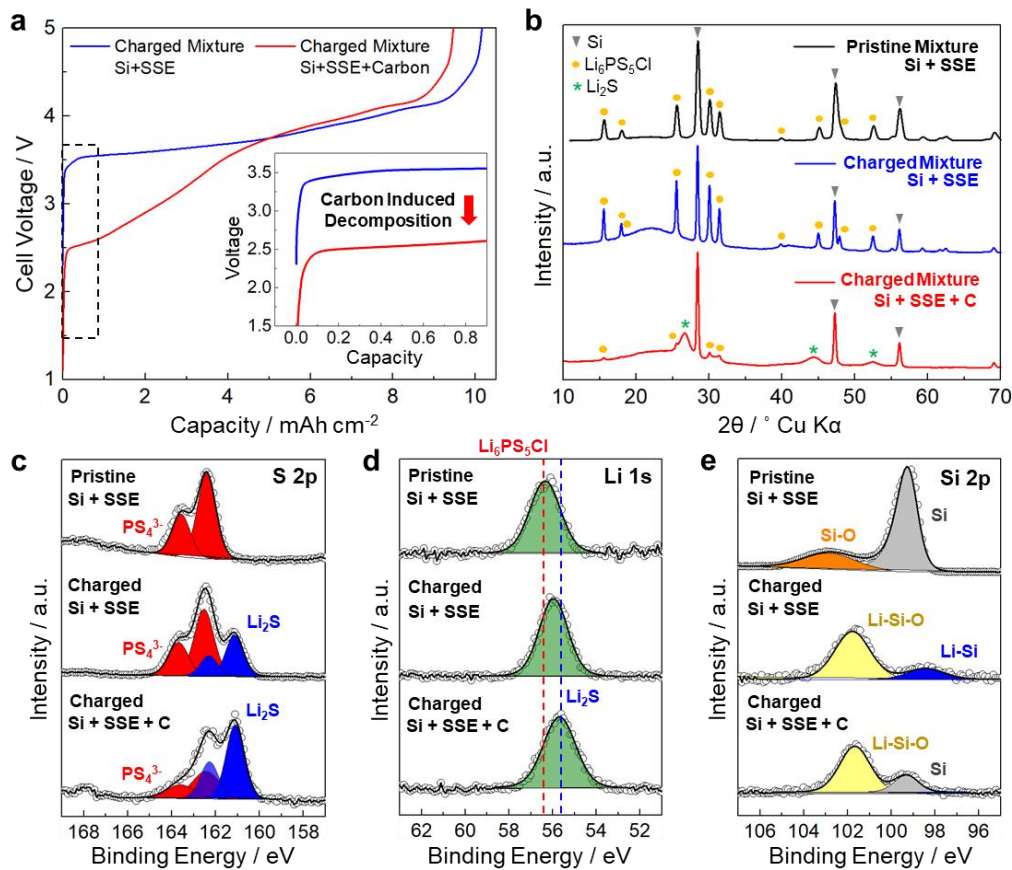

*Fig. 2. Carbon effects on SSE decomposition. (a) Voltage profiles of μSi || SSE || NCM811 cells with and without carbon additives (20 wt.%), inset shows a lower initial plateau indicating SSE decomposition to form SEI. (b) X-ray Diffraction (XRD) patterns, and (c-e) X-ray Photoelectron Spectroscopy (XPS) spectra of the (c) S 2p, (d) Li 1s and (e) Si 2p core regions, showing increased growth of $Li_2S$ interfacial products when carbon additives are used.*

These observations agree with the XPS analysis in **Fig 2c**, where the presence of carbon results in a greater extent of SSE decomposition, as seen by the formation of $Li_2S$ (161 eV) in the S 2p region. Consequently, a larger decrease in peak intensities for the $PS_4^{3-}$ thiophosphate unit signals is observed for the electrode containing carbon (**Fig 2c bottom**) compared to the electrode without carbon (**Fig 2c middle**). Although the Li 1s region (**Fig 2d**) is difficult to deconvolute due to the presence of multiple Li$^+$ species, a shift toward lower binding energies is observed as a result of reduction of Li$^+$ from the pristine SSE. While a smaller shift is observed in the sample without carbon (**Fig 2d middle**), the peak position for the sample containing carbon is dominated by the Li$_2$S signals at about 55.6 eV (**Fig 2d bottom**), reaffirming the previous observation using XRD. In the Si 2p region, a native oxide layer is detected near the surface of the Si particles (**Fig 2e top**). Upon cell charging, this signal shifts to a lower binding energy as a result of lithiation. Interestingly, a peak with a binding energy consistent with Li-Si is found in the sample without carbon, while Si appears to remain unreacted in the sample with carbon. This is likely due to formation of the Li$^+$ consuming SEI products, severely limiting the lithiation of the µSi electrode itself. These results demonstrate the importance of eliminating carbon conductive additives in the µSi electrode when SSEs are used. With the low-density carbon eliminated from the electrode composite, binder (PVDF) use can be minimized to 0.1 wt% of the anode. This amount was found to be ample to prepare slurries for casting of 99.9 wt% µSi anodes, as shown in **Fig S1**.

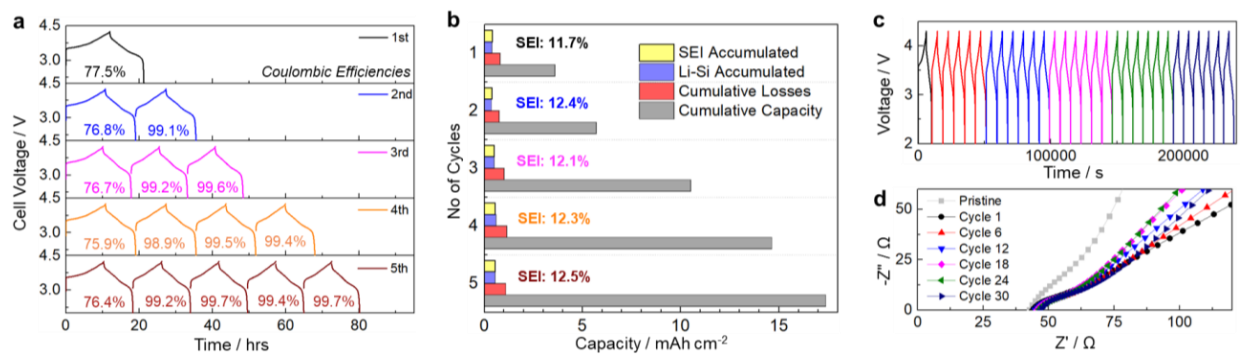

*Fig. 3.* (a) Voltage profiles of full cells used in titration gas chromatography (TGC) experiments. (b) Quantified Li-Si and SEI amounts relative to cell capacity. (c) Voltage profile of full cell used for EIS, and (d) Nyquist plots for different cycle numbers showing limited impedance growth.

*Quantification of SEI Components*. While capacity fade during cycling can be detected as a function of coulombic efficiency (CE%) loss, it is difficult to accurately deconvolute contributions from the SEI or trapped Li-Si respectively. In previous studies, titration gas chromatography (TGC) has been effectively employed to quantify SEI and dead Li growth in Li metal batteries.(*33*) Here, with a similar operation principle TGC is applied to quantify SEI growth and ascertain its passivating and stable nature (**Fig S2**). Five µSi∥SSE∥NCM811 full cells were assembled and cycled from 1 to 5 cycles respectively (**Fig 3a**). After cycling, the remaining Li-Si anode was fully reacted with anhydrous ethanol to generate H$_2$ gas. The H$_2$ gas was then extracted and quantified using the TGC method, allowing the quantification of the active Li$^+$ present in the cycled sample. The difference between the CE% losses and the quantified Li$^+$ allows derivation of the SEI formed. The amounts of SEI accumulated, active Li$^+$ from Li-Si, sum of cumulative losses, and total cumulative capacities are plotted in **Fig 3b**. After the 1$^{st}$ cycle, the total amount of SEI formed was found to be 11.7% of the cell's capacity, and this amount increases slightly to 12.4% in the 2$^{nd}$ cycle. In the subsequent cycles, both the accumulated SEI as well as the active Li$^+$ were found to remain stable and relatively unchanged, indicating interface passivation that prevents unwanted continuous reaction between Li-Si and the electrolyte. As the formation of SEI also results in impedance growth, electrochemical impedance spectroscopy (EIS) was conducted on the full cell over 30 cycles. The voltage profiles are shown in **Fig 3c**. From the pristine state in **Fig 3d**, an initial increase in impedance is observed after the 1st cycle due to the initial SEI formation, an observation in agreement with the TGC measurements. The impedance then remains stable over the subsequent 30 cycles, indicating that the SEI is no longer growing after the first cycle. This passivating nature of the SEI may potentially extend the calendar life of Si based anodes, as shown in self-discharge tests detailed in **Fig S3 & S4**.

*Morphological Evolution*. As illustrated in **Fig 1**, the interfacial contact area in the Si-ASSB cell configuration only occurs along a 2D plane between the SSE layer and the 99.9 wt% µSi electrode. Unlike liquid electrolyte-based cells, where the liquid infiltrates the pores of the electrode and form a SEI that encompasses each µSi particle, µSi particles in the SSE cell remain in direct contact with each other. This allows for fast diffusion of Li$^+$ and transport of e$^-$ throughout the electrode, unhindered by any electronically insulative components such as SEI or electrolyte.

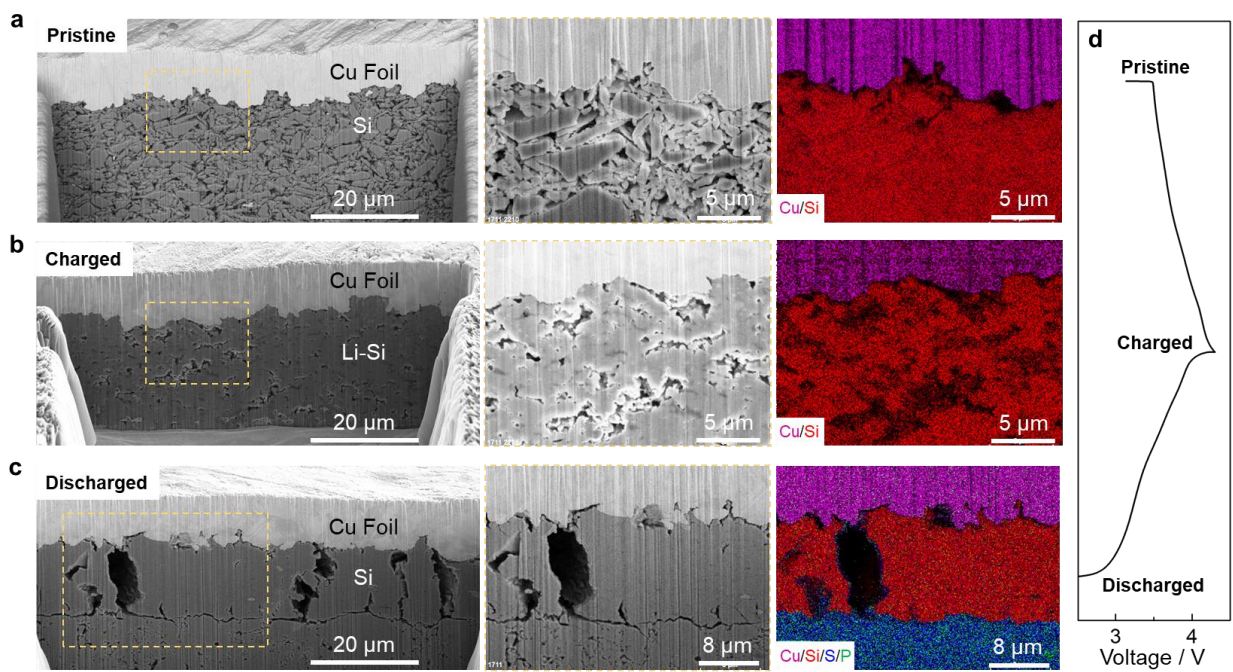

*Fig. 4. Visualizing lithiation and delithiation mechanism of 99.9 wt% Si anodes using focused ion beam (FIB) and energy-dispersive X-ray spectroscopy (EDS) imaging. (a) Pristine porous microstructure of µSi electrode. (b) Charged state with densified interconnected Li-Si structure. (c) Discharged state with void formation between large dense Si particles. Yellow dotted box represents enlarged porous regions of interest for each sample. (d) Reference voltage profile of the full cell charge and discharge.*

**Figure 4** shows a cross-section scanning electron microscopy (SEM) image prepared by focused ion beam (FIB) at the pristine, charged and discharged states of the µSi||SSE||NCM811 full cell. At the pristine state in **Fig 4a**, discreet µSi particles are observed, with sizes ranging between 2 and 5 µm. Despite the initial calendaring pressure of 370 MPa used, the electrode exhibits porosity of about 40% due to the high yield strength of elemental Si that prevents deformation during calendaring. After lithiation and volume expansion in **Fig 4b**, the electrode becomes densified, with the majority of pores disappearing between the initial pristine µSi particles. More importantly, the boundaries between separate µSi particles have entirely vanished. An enlarged view of the more porous region shows that the entire electrode has become an interconnected densified Li-Si alloy. After delithiation and volume contraction (**Fig 4c**), the µSi electrode did not revert to its original discreet micro-particle structure but instead forms large particles between the SSE layer and the copper current collector. Large voids are also formed between these particles. Energy-dispersive X-ray (EDS) imaging confirms that the pores are indeed voids, with no evidence of SEI or SSE present between each delithiated particle. A charge and discharge voltage profile is displayed in **Fig 4d** for reference. The morphological behavior observed is a stark contrast to morphological changes of µSi particles in liquid electrolyte systems (**Fig S6**).

This unique chemo-mechanical behavior of the Li-Si alloy has been previously reported in literature studies on porous Si thin film ASSBs using sulfide SSEs as well, where initial porosity incorporated into the pristine Si thin film electrode was found to be able to accommodate volume expansion during lithiation to some extent.(*11, 23*) However, given the large volume expansion expected in µSi particles (>300%), the 40% porosity found in the pristine electrode is insufficient to fully accommodate this expansion, and thickness changes along the z-axis are still expected. To visualize thickness growth as well as to determine porosity changes during lithiation and delithiation, µSi electrodes with mass loadings of ~3.8 mg cm$^{-2}$ were used in full cells with N/P ratio of 1.1 and their thickness changes during cycling measured using the SEM cross-section. At the pristine state, a thickness of ~27 µm was measured (**Fig S5a**), and after lithiation to Li$_{3.35}$Si, the thickness increased to ~55 µm (**Fig S5b**). This increase falls short of the expected >300% growth(*1*), indicating that a significant decrease in porosity must occur. To rationalize this, expected thicknesses vs porosities were calculated in **Table S1**, which shows a low resulting porosity (<10%) of the ~55 µm µSi electrode after lithiation. This agrees with the observations made in **Fig 4**, where significant densification is observed compared to the pristine state. After delithiation (**Fig S5c**), a thickness of ~40 µm was measured, with a porosity of ~30% calculated using the same approach. The lower porosity at the delithiated state compared to the pristine 40% is expected as some Li$^+$ remains in the anode of the full cell (**Fig 3b**). Despite the relatively large thickness and porosity changes of the anode during cycling, the lithiation and delithiation was found to be highly reversible over a wide range of conditions as discussed in the subsequent section. This suggests that the mechanical properties of the Li-Si and SSE have a crucial role in maintaining the integrity of the interfaces as well as retaining conformal contact with the anode. While contact losses are less likely during lithiation, where volume expansion occurs, it is an important consideration during delithiation. However, despite volume shrinkage, good contact is still maintained between the SSE layer and the porous structure of the delithiated Li-Si (**Fig 4c**). This indicates that some degree of Li-Si deformation occurred during cell cycling under a uniaxial stack pressure of 50 MPa used. While pristine µSi did not deform even under large pressures of 370 MPa, existing reports found that both young modulus as well as hardness of Li-Si alloys decreases significantly as a function of lithiation, with a modulus reaching as low as 12 GPa for Li$_{3.75}$Si, resulting in mechanical properties more similar to metallic Li (modulus of 8 GPa) than pristine µSi (modulus of 92 GPa).(*34-37*) Given that metallic Li was found to achieve sufficient deformation and mechanical contact with SSEs under stack pressures of 5 to 7 MPa(*32, 38*), it can be expected that Li-Si also undergo sufficient deformation during cell cycling under 50 MPa, maintaining the conformal contact with the SSEs used here. Naturally, as pressure is only applied in the z-axis, volume shrinkage would result in pore formation in directions parallel to the electrode during delithiation, an observation also reported in other Si thin film studies.(*11*) As the interfacial contact is maintained along the 2D plane between the SSE and the anode (**Fig 1**), there is no generation of new anode surfaces exposed to the SSE where further SEI grows, enabling high reversibility of the lithiation and delithiation process.

*Electrochemical Performance*. To test the 99.9 wt% µSi in full cells, a high loading NCM811 cathode was prepared using a dry electrode process (**Fig S7**), where Polytetrafluoroethylene (PTFE) was used as a binder to achieve thick electrodes. These electrodes were characterized using X-ray computed tomography experiments and compared with slurry casted ones (**Fig S8**). Unlike conventional slurry cast electrodes, the dry process is able to achieve improved electrode homogeneity as well as higher packing density even with thick electrodes (**Fig S9**). The dry processed cathode composites were then paired against the µSi anodes with an N/P ratio of 1.1 in full cells. While higher N/P ratios reduced the likelihood of Li plating and cell short, it was found to deliver lower average CE% compared to lower N/P ratios (**Fig S10a & b**), thus N/P ratios were kept at 1.1 for all cells.

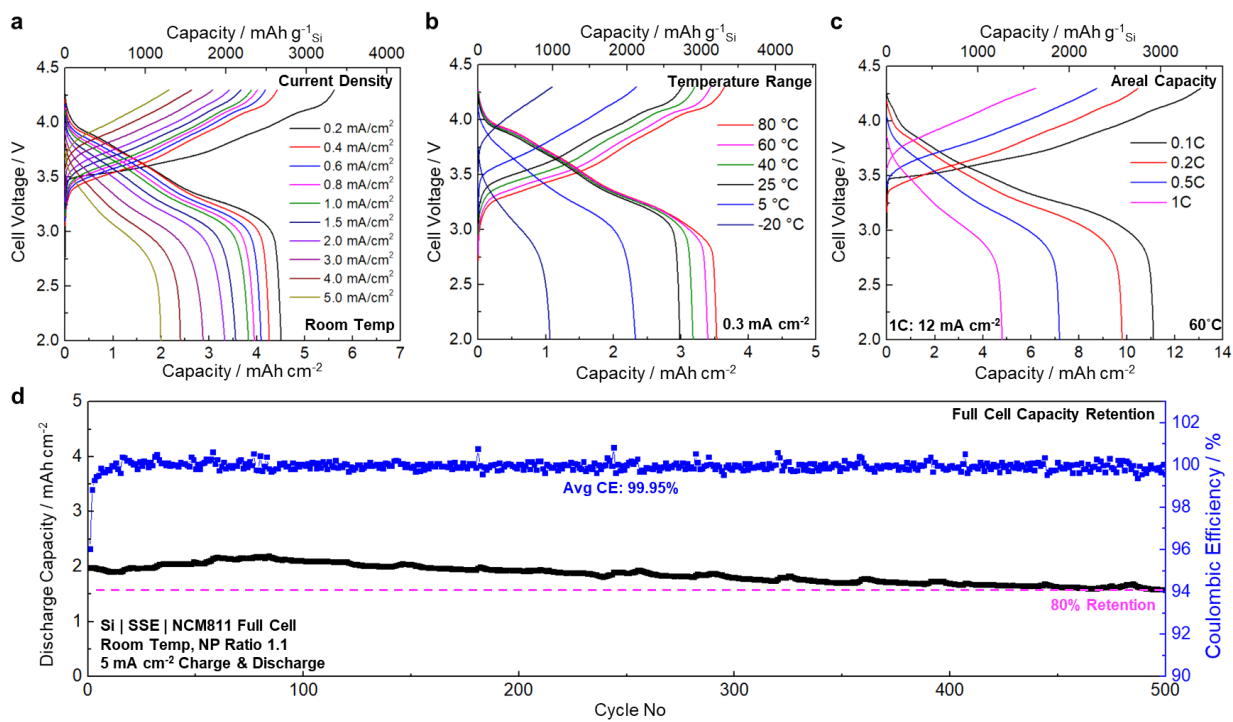

*Fig. 5.* µSi‖SSE‖NCM811 full cell performance: (a) High current densities at room temperature, (b) Wide operating temperature ranges at 0.3 mA cm$^{-2}$, (c) High areal capacities, (d) Cycle life at room temperature, demonstrating overall robustness of ASSB using 99.9 wt% µSi anodes. All cells were tested under similar conditions for charge and discharge between 2.0-4.3V, and an NP ratio of 1.1, based on Si & NCM811 theoretical capacity of 3500 & 200 mAh g$^{-1}$ respectively.

**Figure 5a** shows the room temperature galvanostatic cycling, where current is gradually increased from 0.2 mA cm$^{-2}$ to 5 mA cm$^{-2}$ for both charge and discharge operation. No evidence of cell short occurs even up to 5 mA cm$^{-2}$ at room temperature, significantly higher than the typical room temperature critical current density of Li metal ASSBs reported in the literature.(*31, 32*) In **Fig 5b**, a full cell was charged and discharged over a temperature range between -20°C to 80°C using a moderate current density of 0.3 mA cm$^{-2}$. As temperature increased, the cell's capacity utilization increased as well. While the cell polarization increases dramatically at lower temperatures, likely due to Li$^+$ diffusion limitations within the SSE, the cell does not exhibit any sign of shorting at temperatures as low as -20°C. From **Fig 5c**, a cell with a cathode sized to 12 mAh cm$^{-2}$ was fabricated in order to evaluate high areal loading µSi electrodes. To overcome the bulk impedance of the thick cathode electrode, the full cell was operated at 60°C to enhance Li$^+$ diffusion kinetics. Under current rates of 0.1C (1.2 mA cm$^{-2}$), the µSi anode was found to deliver reversible capacities of more than 11 mAh cm$^{-2}$ with specific capacity of 2890 mAh g$^{-1}$. Under continuous cycling at 1C (12 mA cm$^{-2}$), the µSi anode delivers stable reversible capacity of more than 4 mAh cm$^{-2}$ with specific capacity of 1050 mAh g$^{-1}$ (**Fig S7**). As room temperature charge and discharge remains to be the ideal condition for ASSB operation, cycle life of the µSi||SSE||NCM811 full cell was evaluated by maintaining a current density of 5 mA cm$^{-2}$ at room temperature (**Fig 5d**). The cell was found to achieve a capacity retention of 80% after 500 cycles and an average coulombic efficiency of 99.95%. This capacity fade likely occurs as a result of cathode impedance growth over time resulting from contact losses and cathode interfacial growth.(*39, 40*) Further improvements can be made to engineer the chemo-mechanical properties of the cathode||SSE interface. Nonetheless, the electrochemical results shown above reaffirm the effectiveness of sulfide SSEs to enable 99.9% µSi anodes, free of carbon, yet capable of operating at high current densities, under a wide temperature range, with high areal loadings as well as with long cycle and calendar life. Overall, this approach offers a promising pathway to address the fundamental interfacial bottleneck in the commercialization of low cost and environmentally benign µSi anodes.

**Acknowledgments:** The authors would like to acknowledge the UCSD Crystallography Facility. This work was performed in part at the San Diego Nanotechnology Infrastructure (SDNI) of UCSD, a member of the National Nanotechnology Coordinated Infrastructure, which is supported by the National Science Foundation (Grant ECCS-1542148). Characterization work was performed in part at the UC Irvine Materials Research Institute (IMRI) using instrumentation funded in part by the National Science Foundation Major Research Instrumentation Program under grant no. CHE-1338173. The authors thank Jeong Beom Lee of LGES for the discussion. **Author contributions:** D.H.S.T., Z.C., and Y.S.M. conceived the ideas. D.H.S.T. and H.Y. designed the experiments and cell configuration. Y-T.C., W.B., B.S., W.L., B.L., S-Y.Ham., B.S. and J.S., performed the XRD, XPS, TGC, FIB-SEM and CT experiments. J-M.D., E.A.W., G.D., H.E.H., H.J.H., and H.J., participated in the scientific discussion and data analysis. D.H.S.T. wrote the manuscript. All authors discussed the results and commented on the manuscript. All authors have approved the final manuscript. **Funding:** This study was financially supported by the LG Chem company through the Battery Innovation Contest (BIC) program. Z.C. acknowledges funding from the start-up fund support from the Jacob School of Engineering at University of California San Diego. Y.S.M. acknowledges the funding support from Zable Endowed Chair Fund. **Competing interests:** A joint patent application on this work has been filed between UC San Diego's Office of innovation and commercialization as well as LG Energy Solution, Ltd. **Data and materials availability:** All data is available in the main text or the supplementary materials.

**Supplementary Information:**

Materials and Methods

Figs. S1 to S10

Tables S1

**Fig. 1.** Schematic of 99.9 wt% μSi electrode in an ASSB full cell. 1) During lithiation, a passivating SEI is formed between the μSi electrode and the SSE, followed by lithiation of μSi particles near the interface. 2) The highly reactive Li-Si then reacts with Si particles within its vicinity. 3) The reaction propagates throughout the electrode, forming a densified Li-Si layer.

**Fig. 2.** Carbon effects on SSE decomposition. (a) Voltage profiles of μSi ∥ SSE ∥ NCM811 cells with and without carbon additives (20 wt.%), inset shows a lower initial plateau indicating SSE decomposition to form SEI. (b) X-ray Diffraction (XRD) patterns, and (c-e) X-ray Photoelectron Spectroscopy (XPS) spectra of the (c) S 2p, (d) Li 1s and (e) Si 2p core regions, showing increased growth of $Li_2S$ interfacial products when carbon additives are used.

**Fig. 3.** (a) Voltage profiles of full cells used in titration gas chromatography (TGC) experiments. (b) Quantified Li-Si and SEI amounts relative to cell capacity. (c) Voltage profile of full cell used for EIS, and (d) Nyquist plots for different cycle numbers showing limited impedance growth.

**Fig. 4.** Visualizing lithiation and delithiation mechanism of 99.9 wt% Si anodes using focused ion beam (FIB) and energy-dispersive X-ray spectroscopy (EDS) imaging. (a) Pristine porous microstructure of μSi electrode. (b) Charged state with densified interconnected Li-Si structure. (c) Discharged state with void formation between large dense Si particles. Yellow dotted box

represents enlarged porous regions of interest for each sample. (d) Reference voltage profile of the full cell charge and discharge.

**Fig. 5.** μSi‖SSE‖NCM811 full cell performance: (a) High current densities at room temperature, (b) Wide operating temperature ranges at 0.3 mA cm$^{-2}$, (c) High areal capacities, (d) Cycle life at room temperature, demonstrating overall robustness of ASSB using 99.9 wt% μSi anodes. All cells were tested under similar conditions for charge and discharge between 2.0-4.3V, and an NP ratio of 1.1, based on Si & NCM811 theoretical capacity of 3500 & 200 mAh g$^{-1}$ respectively.